# Effects of PLGA coating on biological and mechanical behaviors of tissue engineering scaffolds

A.M. Maadani, F. Davoodian, E. Salahinejad*

Faculty of Materials Science and Engineering, K. N. Toosi University of Technology, Tehran, Iran

**Abstract**

Scaffolds have a key role in the clinical success of tissue engineering for the regeneration of damaged tissues. Their bio-performance is often described as the extent to which they can provide an extracellular matrix-like environment for cells embedded where their function and growth can effectively continue. For this purpose, tissue engineering scaffolds should exhibit biodegradability, biocompatibility, bioactivity, delivery, and mechanical performance. The use of polymer coatings, especially poly(lactic-co-glycolic acid) (PLGA), on tissue engineering scaffolds has been found to be one of the most effective methods to improve the scaffold properties. This paper reviews the techniques used to coat tissue engineering scaffolds with PLGA and its effects on the mechanical characteristics, biodegradability, biocompatibility, Molecular delivery, and osteointegration of the scaffolds. It is concluded that apart from apatite-formation ability, all bio-functionalities can be tuned through PLGA coatings. This reflects the great potential of this modification approach to be used in tissue regeneration and therapeutic delivery applications.

* Corresponding Author: Email Address: <salahinejad@kntu.ac.ir>

**Table of Content**

**1. Introduction**

With the surge of transplant operations in recent years, the medical industry faced with a serious challenge of inequality between the number of patients and donors. Different solutions have been sought and among those, advances in tissue engineering have allowed it to emerge as a promising discipline through which medical interventions like the accelerated reconstruction of damaged tissues are becoming more and more feasible [1, 2]. The centerpiece of the tissue engineering approach is the scaffold acting as a vital support and guiding platform for embedded cells.

Specific properties are required for tissue engineering scaffolds, including biocompatibility, degradability, osteointegration, the ability to deliver growth factor or other therapeutic agents loaded in them, and appropriate mechanical behavior. A significant purpose of the scaffolds is to enable cells to attach, differentiate, proliferate, and form a tissue that replaces the degrading scaffold; therefore, most scaffolds are not intended for long-term implantations. Consequently, utilizing biodegradable materials in the fabrication of scaffolds and tuning their properties to achieve a desirable balance between the scaffold degradability and the growth rate of the new tissues are highly coveted in the relevant research [3, 4]. Also, when it comes for scaffolds to make chemical bonds to the targeted tissue and promote better stability, bioactivity should be taken notice [5]. A capacity for controlled delivery behavior is also critical for scaffolds employed as carriers since the too slow or too rapid release of the loaded content can lead to inefficacy or toxicity responses, respectively [6, 7]. Biocompatibility is the fundamental property expected from scaffolds since they should allow the seeded cells to anchor to them and provide a suitable environment that biochemically and physically cues differentiation and proliferation. Biocompatibility is characterized by not inducing any long-term inflammatory reactions of the surrounding tissue and the body hosting the implanted biomaterial that cause tissue repair interruptions or in worse cases, rejection [5, 8]. In addition, the mechanical behaviors of the scaffolds should match those of the target organ to prevent the stress shielding effect and protect the embedded cells from mechanical damage [9]. Thus, the main challenge in the progress of tissue engineering scaffolds is to realize a combination of the properties mentioned above simultaneously. To address the challenge, two approaches have been widely researched mainly in the context of bioceramic scaffolds, which are the addition of dopants to scaffolds or the use of biodegradable polymer infiltration/ coatings [10-13].

In the literature, polymer coatings have been numerously reported as a viable modification technique to tune the bio-function of tissue engineering scaffolds, especially bioceramic ones that predominantly suffer from insufficient toughness. In this regard, an assortment of biodegradable synthetic and natural polymeric biomaterials, such as polycaprolactone (PCL), poly (lactide) (PLA), poly (lactic-co-glycolic acid) (PLGA), alginate, gelatin, and chitosan has been studied to act as coatings or infiltered polymers, each augmenting the scaffold performance according to the basic polymer properties and scaffold-polymer interactions [14-16]. Among them, PLGA provides key advantages; principally, the possibility to tune hydrophilicity through the change in the ratio of the monomers, which is vital when biodegradability adjustment is required. Further, the pH buffering impact that PLGA provides during its dissolution in the biological environment can be exploited in the service of improving scaffold biocompatibility [9, 17]. Apart from a comparative mini-review on PLA and PLGA coatings [18], no review paper has comprehensively focused on the influence of PLGA coatings on the bio-functionality of tissue engineering scaffolds, to our knowledge, which is the subject of this paper.

## 2. PLGA as a biomaterial

The most prominent biodegradable polymers with a non-natural source are aliphatic polyesters with adequate biocompatibility and biodegradability. The natural metabolization of their low molecular weight degradation products leads to relatively rapid removal and excretion through the kidneys. The most common monomers used for the production of biomedical aliphatic polyesters are ε-caprolactone (CL), glycolide (GL), and lactide (LA), introducing poly(ε-caprolactone) (PCL), polyglycolic acid (PGA), polylactic acid (PLA), and poly(lactic-co-glycolic acid) (PLGA) as attractive biomaterials [19, 20].

PGA, PLA, and PCL are biodegradable, thermoplastic polymer chains usually produced through ring-opening polymerization [21]. PGA is a hydrophilic, rapidly degrading, and relatively high-strength polyester that makes it an appropriate substance for short-term tissue engineering like absorbable sutures and meshes. Having a semi-crystalline and hydrophilic structure, water penetrates into amorphous zones and dissolves the non-crystalline part of the network through the hydrolytic cleavage of ester bonds, exposing the remaining ester bonds in the crystalline region to hydrolytic attack [22, 23]. PLA is another hydrophobic aliphatic polyester with a degradation mechanism the same as PGA, where water first diffuses into the amorphous region leading to hydrolytic chain scission. The remaining carboxylic groups play a catalytic role giving rise to the heterogenous continuation of degradation, where faster degradation occurs inside the polymer molecule compared to the surface [24, 25]. PLA embodies outstanding behaviors, such as suitable biocompatibility and slow biodegradability, which make it an ideal candidate for producing implants in the forms of plates, pins, screws, rods, anchors, mesh, and delivery systems [26]. PCL, the other standard polyester with a strong hydrophobic nature, is of interest as it can be achieved from inexpensive materials, has a high solubility in organic solvents, and has an excellent ability to create blends with other polymers. This polymer is also degraded by the hydrolysis of ester linkages in biological media [27, 28]. Owing to the low degradability of PCL, it has been researched as a substance for long-term implantable devices, tissue engineering, and slow delivery applications [29].

PLGA is one of the most attractive polymer candidates for tissue reconstruction applications, which is produced via the copolymerization of its monomers, i.e., glycolic and lactic acids. Depending on the ratio of the monomers in the synthesized polymer network, various forms of PLGA can be achieved (Fig. 1). The copolymer shows proper biocompatibility, biodegradability, delivery behavior, and a glass transition temperature from 40 °C to 60 °C. The degradation of PLGA happens by the autocatalytic hydrolysis of ester linkages [30, 31]. It has been also indicated that the period required for the biodegradation of PLGA is associated with the ratio of the monomers utilized: the higher the level of glycolide in the polymer network, the higher the degradability compared to lactides [32]. This flexibility in tuning degradability aligning with excellent biocompatibility makes this polymer suitable for different medical applications like grafts, sutures, surgical sealant films, prosthetic devices, pins, implants, and micro/nano drug carriers [33].

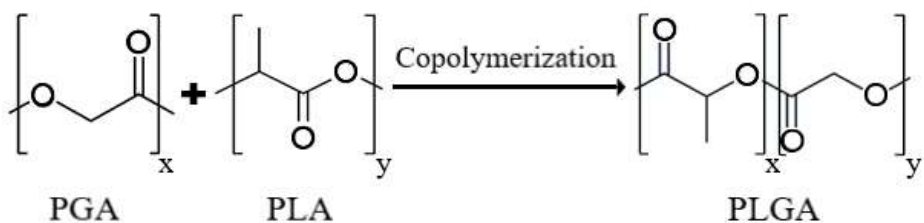

Fig. 1. Synthesis of PLGA through co-polymerization of PGA and PLA.

## 3. Substrate materials used under PLGA coatings

The suitable biocompatibility of bioceramics, as primary substrates for PLGA, is not the only requirement for successful bone tissue engineering scaffolds. Mechanical properties of the scaffolds are also of great importance, especially in load-bearing scaffolds. The major disadvantage of bioceramic scaffolds is their high brittleness and low strength, not allowing them to be used in load-bearing applications alone. To address such weaknesses, coating or infiltrating the scaffolds with PLGA provides an optimal combination of the properties of each phase (structural and mechanical behaviors, osteogenesis and degradability). PLGA-coated bioceramics include hydroxyapatite (HA), tricalcium phosphate (TCP), 58S bioactive glass, $CaSiO_3$, bredigite ($MgCa_7Si_4O_{16}$), diopside ($MgCaSi_2O_6$), and akermanite ($MgCa_2Si_2O_7$). PLGA coating of bioceramic scaffolds has been indicated to toughen and strengthen the structure remarkably [34, 35].

Biometals like titanium alloys are the second candidate to be considered under PLGA coatings. Indeed, metallic implants often are prone to low corrosion resistance in contact with the physiological environment. Thus, untreated metals are rarely used for permanent clinical applications as the release of corrosion ions induces the dysfunction of the adjacent living cells. Nevertheless, as the surface modification of metallic scaffolds, introducing PLGA films not only reinforces the corrosion resistance, but also provides a more favorable tissue-implant interface [36, 37].

There are little research regarding PLGA deposits on polymeric substrates like PCL and silk, but among those available, PLGA is coated mainly to ameliorate the biocompatibility of the construct. It is concluded that all the three types of materials, including ceramics, metals and polymers can be coated with PLGA to enhance their bio-performance. In the following chapters, PLGA coatings deposited on different substrate materials and their consequent improvements are discussed in detail.

## 4. PLGA coating techniques

The surface preparation of the substrate before coating is a critical step to meet the highest performance of coating systems, especially for the coating adhesion to the substrate. Despite the considerable number of articles in the literature on PLGA-coated scaffolds, the surface preparation of the scaffolds has not been focused so far. Given other polymer coatings deposited on tissue engineering scaffolds, it is concluded that surface treatments generating functional groups at the interface of the two materials affect the coating adhesion. For instance, the chemical modification of polymer scaffolds has been performed by reagents like acids and oxidizers, which can increase the surface polarity, molecular forces between the coating and substrate, and consequently the adhesive strength. Plasma treatment is another effective way of improving the inherently poor surface properties of substrates without changing the overall bulk characteristics [38, 39].

There are two typical approaches to coat scaffolds with PLGA; dip-coating and electrospinning. Dip-coating is a simple and effective method to deposit soluble polymers, including PLGA on dense and porous substrates. According to Fig. 2a, this method involves immersing the scaffold into a solution of PLGA with a particular concentration, where a higher polymer content leads to a thicker coating after solvent evaporation. The coating characteristics is a function of dip-coating parameters, including polymer concentration, solvent composition, number of dip-coating and drying passes, immersion duration, and deposition temperature [40, 41]. However, there are drawbacks regarding the scaffold coating uniformity through simple dip-coating, where vacuum and centrifuge-assisted setups can be employed to achieve better homogeneity. Exerting vacuum on a dip-coating polymer setup can lead to a fuller penetration of the solution into the porous substrate, which makes it possible to obtain a porous scaffold encapsulated in the polymer [42, 43]. Centrifuging the scaffolds immediately after immersion in the polymer solution has been also employed to remove the excess solution. This method yields a substrate with a better polymer coverage of struts and pores, which is critical especially for mechanical reinforcement of the ceramic scaffolds by coated polymers [9, 44].

Electrospinning is another approach used to apply PLGA coating on tissue engineering scaffolds. This route consists of dissolving the polymer in a solvent and then imposing a strong electric potential to the solution to make the droplets of the solution accelerate toward the target and induce a spiral motion in them. Afterwards, the solvent is evaporated from the polymer solution and a dry PLGA coating is deposited on the targeted scaffold (Fig. 2b). In an electrospinning apparatus, a capillary tube, a high-voltage supply, and a collector made of a conducting substance are necessary components. In addition, numerous parameters like temperature, pressure, tip-to-target distance, diameter of the needle, operating voltage, and flow rate are influential in the electrospinning process [45, 46].

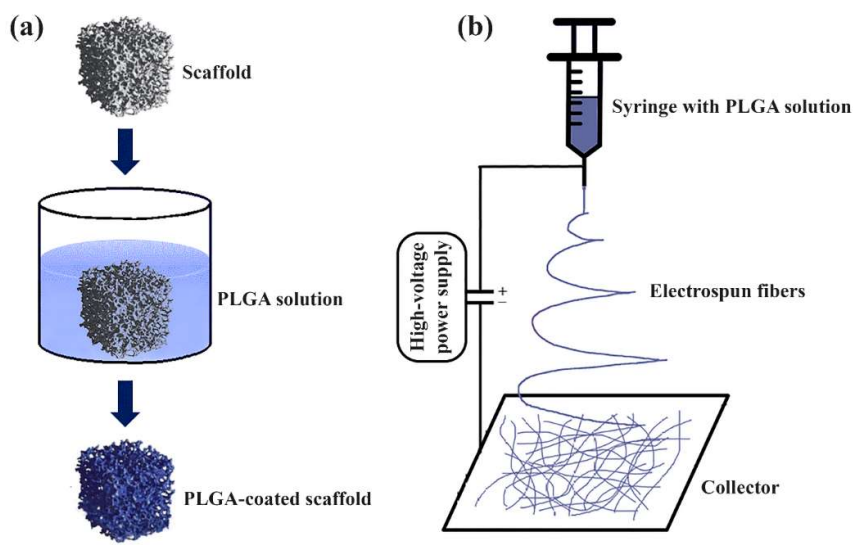

Fig. 2. Setup of dip coating (a) and electrospinning (b).

Comparing these two main methods, dip-coating is a straightforward, inexpensive, and versatile route by which the thickness of the deposited layer can be easily adjusted within the micron accuracy. It is also suitable for complex-shaped and curved parts that cannot be coated by other methods. But its main drawback is the inability to achieve an even thickness of the film across the substrate surface or the coating buildup that occurs as the material is drip-dried. Electrospinning is a versatile method by which ultrathin coatings within the nanometric range with very high specific surface area and porosity levels can be made; however, the route is costly and complex. It is worth noting that the polymer solution concentration is a limiting factor in electrospinning as a stable polymer coating is only achieved when the viscosity is high enough [47, 48]. Mainly due to the limited research works performed on electrospun PLGA coatings, there is no data to compare the mechanical and biological properties of scaffolds coated with PLGA by the spin-coating and electrospinning methods. However, Kelvin et al. [49] compared characteristics of filter papers modified with PCL electrospinning and dip-coating. It was found that dip-coating with PCL reduces the porosity of the scaffolds, thus lowering existing spots for cells to anchor and restricting the penetration and scattering of the cells within the structure. But the electrospun-coated samples enjoyed better cell distribution and attachment as well as higher mechanical performance with better adhesive strength at the polymer-scaffold interface owing to the induced favorable surface morphology and higher porosity. Hence, it is expected that electrospun PLGA coatings introduce superior biological and mechanical properties in comparison to the other coating method.

Adhesion of coatings depends on chemical bonds, surface free energy, cleanness, roughness, and mechanical properties [39, 50]. Although the mechanism of adhesion has not been evaluated for PLGA to date, it can be expected that the adhesion mechanisms similar to other polymer coatings are involved, including mechanical coupling and molecular bonding. Mechanical coupling consists of the interlocking of the coating layer into the irregularities of the substrate surface, in combination with the formation of van der Waals bonds between the two surfaces. Molecular bonding, which is the most widely accepted mechanism, requires the generation of intermolecular forces like covalent bonding between the coating and substrate [51, 52]. It is also worth mentioning that the adhesive strength of the applied coating can be enhanced through some subsequent operations. For instance, Zhao et al. [53] demonstrated that the following centrifugation of coated samples right after the dip-coating process can be consequential in reaching higher polymer-substrate adhesion due to the generation of more uniform coverage.

## 5. Effects of PLGA coating on properties of scaffolds

PLGA can be used to alter the bio-performance of tissue engineering scaffolds by virtue of its controllable biodegradability, desirable drug release behavior, proper biocompatibility, and stable mechanical characteristics. A good number of reports have been published on the properties of PLGA-coated scaffolds, which will be discussed in this chapter.

### *5.1. Mechanical properties*

Among specific requirements that should be met in the manufacturing of bone tissue engineering scaffolds for biomedical applications, mechanical properties should be highly regarded. For instance, the significance of the mechanical properties of regenerative scaffolds in retaining the required space for tissue reforming is undeniable since the osteogenesis of bone tissue engineering scaffolds is associated with structural stability [54, 55]. In this regard, tensile and compressive tests are common to assess scaffolds mechanically. As they should have mechanical behaviors aligned with those of the adjacent tissue, different ranges of mechanical properties are needed depending on the scaffold applications [56]. To alter the mechanical characteristics of scaffolds, coating them with polymers has been shown as a potential approach in many studies, among which PLGA has been given a high deal of attention owing to its adjustable bio-performance, mechanical properties, and infiltration.

Khojasteh et al. [57] performed compression tests to assess the strength of bare and PLGA-coated β–TCP scaffolds at various PLGA concentrations and thicknesses. As concluded, the samples coated with the PLGA concentrations of 5, 10, and 20% possess higher compressive strength (3.10, 4.95, and 6.62 MPa, respectively) than the bare counterpart (0.42 MPa). That is, the PLGA coating offers compressive strength near the strength of the human cancellous bone (2-10 MPa [58]) since the coating fills crack-like defects. Furthermore, by enhancing the PLGA coating concentration, the more fraction of the scaffold is coated and the struts become thicker, improving further the mechanical behaviors of the scaffolds. In this research, the toughness of the scaffolds was also measured, revealing that as the coating concentration increases from 5 % to 10 and 20%, the toughness goes up further from 0.94 to 1.73 and 2.18 MPa, respectively, which all are notably more than that of the bare specimen (0.06 MPa).

Kang et al. [42] investigated the mechanical characteristics of bare and PLGA-coated β-TCP scaffolds. The compressive strength of the coated sample was 4.19 MPa, whereas it was 2.90 MPa for the bare one; moreover, the bending strength increased from 1.46 MPa to 2.41 MPa as a result of coating. It was believed that the PLGA film penetrates into some defects in the struts, leading to the remarkable enhancement in the compressive and bending strengths of the coated samples, which is aligned with the strength of the cancellous bone. Furthermore, the results demonstrated that the applied polymer coating reinforces the toughness of the scaffolds significantly from 0.17 MPa to 1.44 MPa. Likewise, Yoshida et al. [59] found that the compressive strength of PLGA-coated β-TCP scaffolds is even six-fold higher than the bare ones. Owing to the reinforcing impact of the PLGA film on the stability of the specimens, the essential space for tissue regeneration is maintained.

Guo et al. [60] studied the reinforcement of mechanical behaviors in HA nanowires by applying different kinds of polymer coatings, including PLLA, PCL, PLGA, PLCL, and gelatin. In this regard, all the samples were immersed in water and their degradation was monitored alongside it. Generally, the mechanical resistance of the scaffolds is reduced throughout soaking time. However, in the early weeks of immersion, all types of the used polymer coatings play a crucial role in maintaining the mechanical behaviors of the HA samples to a different extent, which is in the order of gelatin, PLLA, PLGA, PLCL, and PCL from the most effective to the most inferior ameliorating influence.

Zhao et al. [43] assessed the effect of PLGA coating with different viscosity levels (2.3, 1.7, and 0.5 dl/g) and concentrations (0, 5, 7.5, 10, 12.5, and 15% w/v) on the mechanical characteristics of $CaSiO_3$ scaffolds. It was found that the compressive strength is enhanced as the PLGA viscosity increases. However, the coatings with the intrinsic viscosity of 1.7 and 0.5 dl/g are considered more desirable since by increasing the viscosity of PLGA to 2.3, a nonuniform coating is achieved and some large pores of the scaffolds are unfavorably blocked. Moreover, it was deduced that the coated samples have better mechanical properties than the bare ones, and those with the more PLGA coating concentrations show more toughness. Comparing the fabricated scaffolds, the samples coated with 12.5% PLGA enjoy the highest compressive strength.

Keihan et al. [61] compared the compressive strength of PLGA-coated bredigite scaffolds with bare ones. It was found that the strength of the samples is enhanced considerably and gets close to that of trabecular bone after coating with PLGA. It was inferred that first, the innate strength of the applied coating, second, the penetration of PLGA into nanopores and some micropores of the scaffolds, and third, the increase in the thickness of the struts as a result of the coating contribute to this improvement. The same findings were pointed out by Jadidi et al. [9].

O'Shea et al. [62] tested the compressive strength of 58S bioactive glass-ceramic scaffolds coated with PLGA. In this study, the compressive strength of the bare samples was 0.12 MPa which was improved to 0.25 MPa thanks to the polymer coating as it enhances the structural integrity of the scaffolds.

Maurmann et al. [45] focused on the mechanical behaviors of PLGA-coated PCL scaffolds. In this research, it was concluded that all of the produced specimens show acceptable mechanical stability, although the polymer coating does not much influence the Young's modulus and strength of the bare PCL samples.

Sahoo et al. [63, 64] also investigated PLGA-coated silk microfibrous scaffolds mechanically. It was declared that mechanically robust samples were attained after coating. Although the slow-degrading knitted silk substrate is responsible for providing the mechanical support needed throughout the healing process, further mechanical properties can be conferred by coating of the samples with linearly aligned nanofibers which are better than random orientation.

Sadiasa et al. [65] evaluated the effect of PLGA/BCP coatings on the compressive strength of BCP/$ZrO_2$ scaffolds. The observed enhancement in their mechanical behaviors owing to the coatings was assigned to the reduction of defect sizes and the ductile characteristics of PLGA. Besides, by increasing the PLGA amount in the film, more mechanical stability is achieved.

Concerning composite scaffolds, Miao et al. [66] measured the compressive modulus and strength of bare and PLGA-coated HA/TCP scaffolds. Given the compression testing results, it was inferred that PLGA fills micropores of the samples and makes the composite struts tougher and stronger, consequently enhancing compressive strength. Although the mechanical behaviors of the fabricated PLGA-coated HA/TCP samples are still below those of the human cancellous bone because of their high porosity level and large pores, they can be improved by increasing the applied coating thickness.

As a result, PLGA coatings can offer considerable alterations in the mechanical behavior of scaffolds, attributed to the reduction of defects, reinforcing of the struts via infiltration, and the ductile nature of this polymeric biomaterial. Moreover, with increasing the PLGA thickness, the mechanical properties are improved further and tend to mimic those of natural tissue.

### *5.2. Biodegradability*

The degradation rate of scaffolds, an essential parameter affecting the tissue regeneration process, can be controlled via polymer coatings. Tissue engineering scaffolds, in particular bioceramic ones, often suffer from inappropriate biodegradability when in contact with the physiological medium. This can increase the concentration of scaffold degradation products at the damaged tissue site, disrupting local pH, and leading to immune responses and the scaffold failure [67]. Also, it is vital for all kinds of scaffolds to maintain their structural integrity while tissue regeneration is underway. In this regard, the utilization of polymer films is a practical approach to adjusting the scaffold degradation rate to the target tissue regeneration rate, among which PLGA is often the optimal candidate by providing suitable coverage on the scaffold struts, alteration to the scaffold macropore sizes, and a barrier against the inward diffusion and direct contact of biological fluids with the scaffold structure [62, 68].

Guo et al. [60] compared the degradation behaviors of HA nanowires coated with biodegradable polymers, including PLGA, PCL, poly(lactide-co-caprolactone) (PLCL), gelatin (GEL), and poly(L-lactide) (PLLA). Given the low resorption rate of HANWs in the study duration, the weight loss of the encapsulated HANW samples is attributed to the hydrolysis of the polymer coatings. After 32 weeks, 100 wt% of the GEL, 94 wt% of PLLA, 97 wt% of PLGA, 80 wt% of PLCL, and 63 wt% of PCL coatings were degraded. It means that the scaffold coated with GEL enjoyed the fastest degradation rate owing to the hydrophilic nature of the polymer; and those coated with PCL and PLCL were degraded very slowly because of the high hydrophobicity of the coating. The PLGA-coated and PLLA-coated samples show a moderate degradation behavior as these polymers are not quite hydrophobic in nature (Fig. 3).

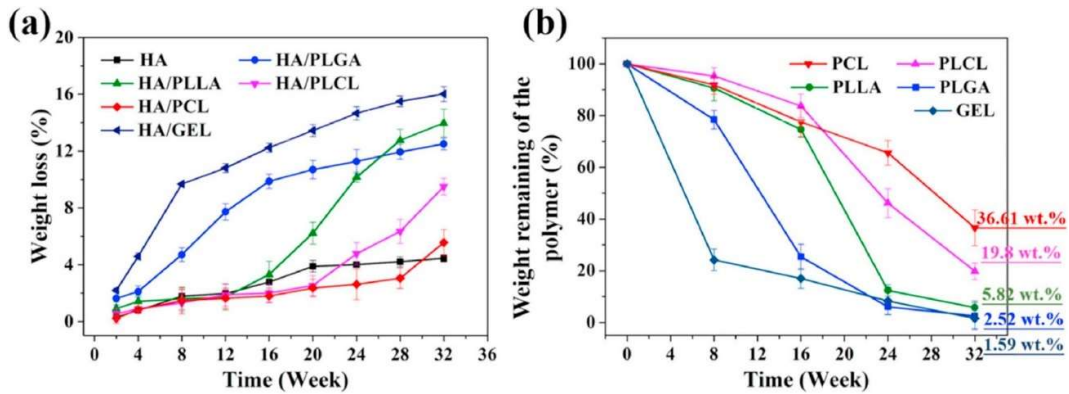

Fig. 3. Wight loss of scaffolds coated with the various polymers (a) and weight remaining of the polymer coatings (b), reprinted with permission from Ref. [60].

Zhao et al. [43, 69] evaluated the biodegradability of bare and PLGA-coated $CaSiO_3$ scaffolds by immersing them in the Tris-HCl buffer solution and measuring pH changes and scaffold weight loss at different times. It was figured out that the pH and weight loss of the bare samples are more than those of the PLGA-coated ones since the coating layer impedes the rapid degradation of $CaSiO_3$ scaffolds. It was also observed that the degradability is declined by increasing the concentration of the PLGA coating.

Keihan et al. [61] investigated the biodegradability of uncoated and PLGA-coated bredigite scaffolds through soaking *in vitro* in the simulated body fluid (SBF). For all the samples, a sharp increase was observed in pH at first owing to the ion exchange of the ceramic cationic ions with $H^+$ or $H_3O^+$ of SBF. The results show that pH for the PLGA-coated samples increases slightly after initiating degradation. Consequently, it was concluded that the PLGA coating prevents the environmental pH from rising too rapidly by controlling the degradation rate. Similar results were drawn by Jadidi et al. [9, 44], where the buffering impact of PLGA is enhanced with the concentration and thereby thickness of PLGA. On the one hand, the release rate of the species from bredigite is confined by increasing in coating thickness acting as a barrier against degradation; on the other hand, the level of acidic products (glycolic and lactic acids) resulting from PLGA degradation and the subsequent buffering effect are increased with the addition of the PLGA concentration.

Sadiasa et al. [65] studied the influence of PLGA/biphasic calcium phosphate (BCP) composite coatings with 33, 50, and 66 %w PLGA content on the biodegradability of BCP/$ZrO_2$ scaffolds. The results revealed that the PLGA/BCP coatings limit the weight loss of the scaffolds during degradation. Furthermore, it was found that by increasing the PLGA amount in the coating, a steeper reduction in the degradation rate is observed for two reasons. First, the higher PLGA content attenuates the porosity of the surface and the penetration of phosphate-buffered saline (BPS) into the substrate. Second, PLGA is a hydrophobic polymer bringing about the lower hydrophilicity and degradability of the scaffolds with more PLGA contents.

In conclusion, PLGA is a promising candidate for coating of scaffolds from the biodegradation viewpoint. This polymer controls the degradation rate of the substrate by, first, reducing the pore size that hinders the rapid penetration of the physiological medium into the structure and, second, covering the scaffolds that prohibits direct contact with the environment. Owing to the amorphous form of this polymer, water molecules diffuse quickly into the structure, which means that PLGA-coated scaffolds do not degrade so slowly.

### *5.3. Biocompatibility*

Biocompatibility, the ability of an implanted matter to function without detrimental local or systemic responses in the body, is an essential criterion for biomaterials and tissue engineering scaffolds, which is determined by various parameters like the structure, morphology, and the composition of scaffolds [70]. Although the medical use of scaffolds has been noted in the literature, there remains an issue of biocompatibility. Nowadays, polymer coating of tissue engineering scaffolds has been investigated in much research as a way to reinforce biocompatibility. Various polymers have been coated on scaffolds, and among them, PLGA is an attractive choice due to its considerable biocompatibility and tunning of the bio-functions of the substrate.

Yoshida et al. [59] used mouse osteoblastic MC3T3-E1 cells to assess the biocompatibility of β-TCP scaffolds coated with PLGA. Despite the acidic by-products of PLGA decomposition, adverse healing effects associated with the polymer were rarely observed, which conveys the low cytotoxicity of the coated specimen.

Khojasteh et al. [57] assessed the biocompatibility of bare and PLGA-coated β-TCP scaffolds with canine mesenchymal stem cells (cMSCs). The cell adhesion, spreading, and proliferation were found to be normal on both the bare and coated samples, although the coated one demonstrates higher biocompatibility.

Kang et al. [42] investigated rat bone marrow stromal cells (rBMSCs) proliferation and differentiation on TCP scaffolds coated with PLGA. As turned out in Fig. 4, cells tend to proliferate and differentiate on both the uncoated and coated specimens. The cell viability was also tested by the MTT method, showing that the coated sample represents slightly a better osteogenic function and cell differentiation as proof of the positive impact of the coating on biocompatibility.

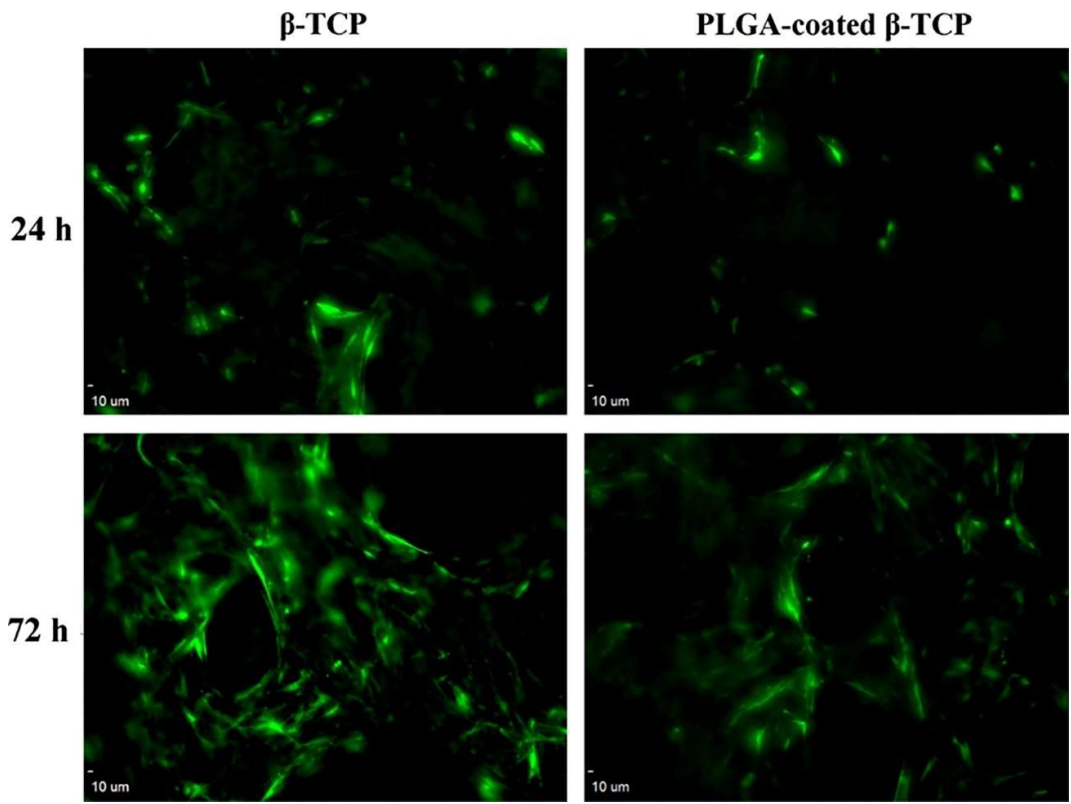

Fig. 4. Confocal laser scanning microscopic graphs of rBMSCs on bare and PLGA-coated TCP scaffolds after different periods, reprinted with permission from Ref. [42].

Zhao et al. [69] studied the biocompatibility of PLGA-coated $CaSiO_3$ scaffolds with BMSCs. The bare $CaSiO_3$ scaffolds show a high degradation rate owing to the existence of many micropores inside, leading to an increase in pH and an adverse effect on the cells metabolism and function. However, coating with PLGA provides a more biocompatible substance for cell spreading, adhesion, and proliferation. Moreover, it was found that the biocompatibility of the PLGA-coated specimens is enhanced after the coating thickness is increased.

Keihan et al. [61] investigated the human osteoblast-like MG-63 cell cytocompatibility of PLGA-coated bredigite scaffolds via the MTT assay. Based on the results, a significant proliferation was reported owing to the presence of the PLGA coating layer, which means that the applied coating supports cellular activities. Also, Jadidi et al. [44] focused on the influence of PLGA coatings with concentrations of 5 and 10% on the biocompatibility of vancomycin-loaded bredigite scaffolds. The bare samples show the lowest dental pulp stem cell viability because of the fast release of vancomycin (Fig. 5) owing to the fast resorption of bredigite. However, the PLGA coatings restrict both the fast drug release and bredigite bioresorption accompanied by enhanced physiological pH, leading to the highest biocompatibility among the assessed specimens. In addition, biocompatibility is further improved with the increase of the PLGA concentration from 5% to 10%.

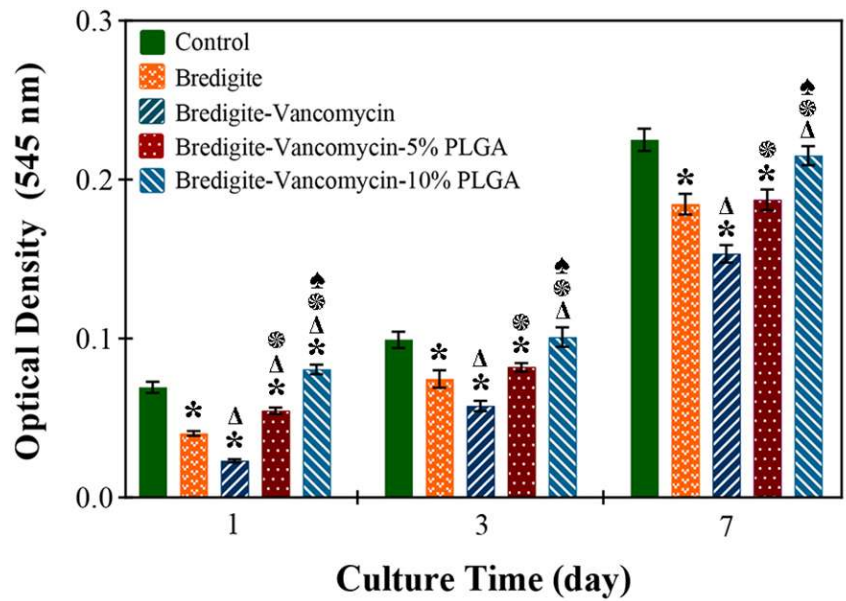

Fig. 5. Impact of PLGA coatings on the dental pulp stem cell cytocompatibility of vancomycin-impregnated scaffolds, reprinted with permission from Ref. [44].

Zirak et al. [71] evaluated the biocompatibility of PLGA-coated vancomycin-loaded Mg-Ca silicate microspheres, including bredigite, diopside, and akermanite. It was found that biodegradability is a dominant factor in determining biocompatibility via bringing about the metabolic alkalosis effect and the fast drug release. The PLGA coating not only limits the explosive drug release but also controls the resorption rate of the ceramics, which leads to higher cell viability on the capsulated carriers. Comparatively, the akermanite substance exhibits the highest cytocompatibility due to a compromise of metabolic alkalosis, drug release, and beneficial (Si, Mg, and Ca) ions release.

O'Shea et al. [62] looked over the biocompatibility of PLGA-coated 58S bioactive glass-ceramic scaffolds via seeding BMSCs. Accordingly, the improved cell migration, attachment, and proliferation in the coated sample were observed.

Respecting polymer-based scaffolds, Maurmann et al. [45] evaluated the impact of PLGA films on the biocompatibility of PCL scaffolds. It was realized that the number of dental pulp stem cells on PLGA-coated PCL samples increases over time. Indeed, the polymer coating gives the samples sufficient mechanical support for cell adhesion and maintains high porosity.

The BMSCs proliferation on silk scaffolds coated with PLGA nanofibers was explored by Sahoo et al. [63]. In this study, on the one hand, well-proliferated cells both on the nanofibrous surfaces and in the depth of the scaffolds, and on the other hand, new extracellular matrix filling pores between the nanofibers were detected. This was mainly assigned to the PLGA film providing a suitable surface for cellular activities and simulating the seeded cell matrixes. Similar results have been found in another research on electrospun PLGA-coated microfibrous silk scaffolds by Sahoo et al. [64], where the PLGA fibrous coating offers a biomimetic platform for cell adhesion. Thus, in the coated construct, better cell proliferation on both the surface of the coating fibers and the inner parts of the scaffold.

Sadiasa et al. [65] used the MTT assay to evaluate the biocompatibility of PLGA/BCP-coated simvastatin-loaded BCP/$ZrO_2$ scaffolds. In this work, a substantial enhancement in the number of MG-63 cells seeded on the samples was seen as a result of the applied polymer coating holding a sustained mode of simvastatin delivery and improving the mechanical properties. This reflects the ability of the BCP/PLGA-coated BCP/ZrO$_2$ scaffolds as a suitable substrate to support cell proliferation.

Miao et al. [66] assessed the effect of PLGA coating on BMSCs attachment on HA/TCP scaffolds through electron microscopic observations. It was indicated that the cells are well adhered to the strut surface of the specimens and penetrate into their structures significantly owing to the interconnected pores in the PLGA-coated HA/TCP specimens and the adequate cytocompatibility of the phosphate and PLGA phases (Fig. 6).

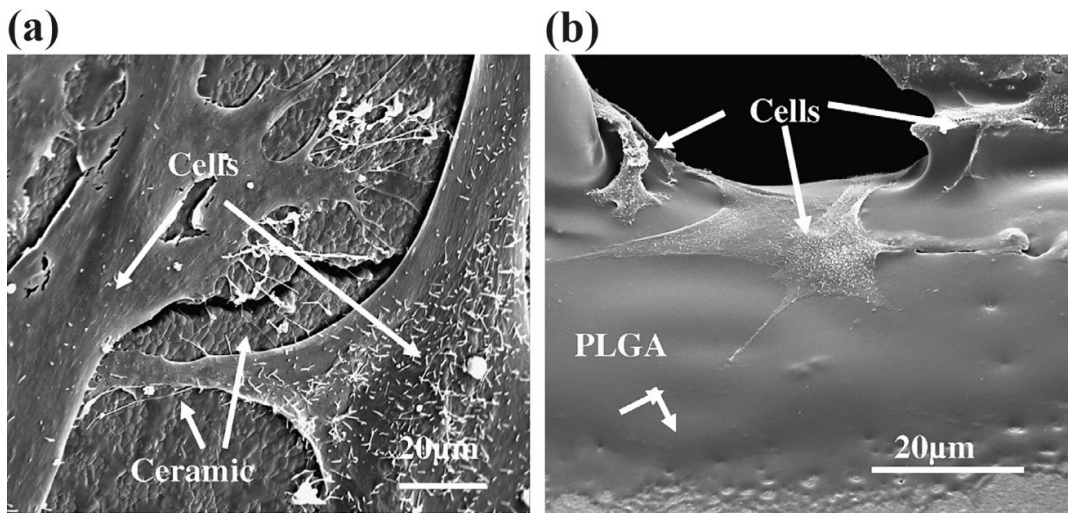

Fig. 6. Microscopic images of BMSCs on bare (a) and PLGA-coated HA/TCP (b) scaffolds, reprinted with permission from Ref. [66].

To sum up, PLGA coatings notably limit the primary burst release of therapeutic agents, slow down the degradation rate of scaffolds, and particularly buffer the local medium alkalized owing to the degradation of bioceramics via the polymer decomposition products. Also, it preserves the suitable interconnectivity of pores in the construction required for (1) proper cellular adhesion, penetration, migration, (2) sufficient supply of nutrients to cells and waste substances out of the porous substances, and (3) gas exchange (i.e., $O_2$ and $CO_2$). They all confirm the desirable biocompatibility of PLGA-coated scaffolds.

### *5.4. Molecular delivery*

Scaffolds are widely used in drug, gene, and growth factor delivery applications; therefore, their delivery performance is of highly importance. On the one hand, provided that the loaded content is secreted in large amounts or rapidly from the carrier, it may cause toxicity, and/or most of it is released before completing treatment. On the other hand, if the delivery takes place in small amounts or at a slow rate not satisfying the therapeutic window, the healing process might not be accomplished. Therefore, scaffolds should show an appropriate delivery behavior according to the loaded compound [72, 73]. However, tissue engineering scaffolds generally do not exhibit an ideal delivery performance. One of the methods commonly employed to make improvements in the delivery kinetics from scaffolds and their loading capacity is coating them with PLGA.

Khojasteh et al. [57] assessed the influence of PLGA coating on the delivery behavior of β-TCP scaffolds loaded with vascular endothelial growth factor (VEGF). It was found that the loading capacity of the bare β-TCP sample is limited due to the poor physical absorption of the drug, but encapsulating VEGF in PLGA coated on the scaffolds is suitable to alter the loading amount. It was also reported that VEGF is released from the coated samples with a primary burst release mode owing to the burst release of the growth factor from the surface and a following sustained release due to the distribution of VEGF in the polymer matrix.

Zhao et al. [43] evaluated the release behavior of $CaSiO_3$ porous scaffolds coated with PLGA after loading with Ibuprofen (IBU) and Rhodamine-B (RB). It was indicated that most of IBUs and RBs are explosively secreted from the bare specimens in the early hours owing to the existing weak Van der Waals bonds between the scaffold surface and drug. On the contrary, the PLGA-capsulated samples offered a more suitable delivery mode of IBUs and RBs for longer times to stop bacterial infection as the coating layer prevented rapid drug elution.

Zirak et al. [71] evaluated the impact of PLGA coating on vancomycin-loaded bredigite, diopside, and akermanite microspheres used against Staphylococcus aureus. The data indicated that the bare scaffolds have a burst release mode in primary periods, followed by a sustained mode of release (Fig. 7a). The PLGA film restricts the initial burst release from all the specimens (Fig. 7b) by acting as a physical barrier for the incorporated drug molecules. Moreover, during the test period, the level of vancomycin released from all the coated samples was above the required level of treatment, showing the capability of inhibiting bacterial growth within the therapeutic window. Also, Jadidi et al. [44] assessed the drug delivery of PLGA-capsulated vancomycin-loaded bredigite scaffolds for osteomyelitis treatment. It was realized that the content of the drug loaded in the bare samples is 66.6%, and 94.7% of it is released during the first nine hours. However, after PLGA coating, the duration of the burst release is declined notably, and a sustained drug delivery is achieved. It is worth noting that the amount of the drug released from the PLGA-capsulated scaffold was desirably beyond the critical level of vancomycin against staphylococcus aureus-caused infection over the whole study time.

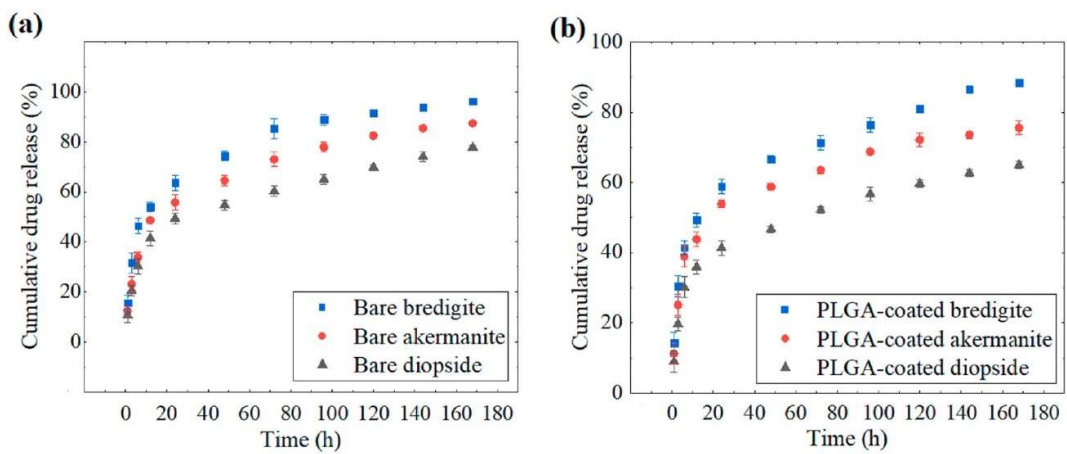

Fig. 7. Level of vancomycin release from the uncoated (a) and PLGA-coated (b) microspheres, reprinted with permission from Ref. [71].

Sahoo et al. [64] developed knitted microfibrous silk scaffolds coated with bioactive basic fibroblast growth factor (bFGF or FGF-2)-releasing ultrafine PLGA fibers. The electrospun PLGA fibers caused a sustained release that activates signaling events in bone marrow stromal cells (BMSCs) and thereby stimulates their differentiation and proliferation. It is accepted that bFGF can be released from electrospun PLGA through a combined mechanism of degradation and diffusion.

Sadiasa et al. [65] focused on the drug delivery behavior of PLGA/BCP-coated simvastatin-loaded BCP/$ZrO_2$ scaffolds. Given the presence of the drug in the coating layer, the drug is released rapidly in initial periods. Nevertheless, after the initial burst release, the drug is delivered in a sustained mode by virtue of the PLGA film as a barrier to control the release. Besides, the medium quickly infiltrates inside the structure of the scaffold in lower concentrations of PLGA as a consequence of the higher porosity existing in the construction, leading to a higher effective release rate.

Briefly, the encapsulation of scaffolds with PLGA can hopefully inhibit the burst release of therapeutic agents, and the release kinetics from PLGA can be satisfactorily tunable. Generally, the thickness of the polymer film is the deciding factor in the adjustment of the scaffold release kinetics, where a higher thickness leads to more prolonged degradation, consequently more difficult diffusion of the loaded agent into the matrix, and the delayed therapeutic release from the coated carrier. Optimizing the PLGA coating thickness is readily attainable through the polymer concertation in the dipping stage and the number of the scaffold drying and immersing cycles. Other factors like the relative hydrophilicity of drug molecules and polymer coating can also affect the release kinetics. The relative hydrophilicity of PLGA is construable, granting even more control over not only the biodegradation rate but also the drug-polymer interactions during the release period. Thus, PLGA coating is regarded as a promising candidate to tune the local delivery kinetics for bone tissue engineering.

### 5.5. Osseointegration and osteogenesis

Bioactivity or osteointegration is a crucial property for implanted biomaterials used in tissue engineering deciding whether the foreign object (i.e., scaffold or implant) bonds to the adjacent living tissue in the body, starting with the ion-exchange controlled formation of apatite on their surface. Such apatite layer can be reproduced artificially in SBF by virtue of ion levels almost similar to the human plasma, which can be a valuable approach to assess the apatite-formation ability of biomaterials and therefore estimate their bioactivity [74, 75].

Yoshida et al. [59] examined PLGA-coated β-TCP scaffolds impregnated with FGF-2 histologically. Findings showed that the open cellular structure of the construct is held; that is, the coated β-TCP scaffold possesses a considerable inner infiltration platform. In this group, bone augmentation was promoted, and the interconnected feature of the scaffold struts was filled with new bone and connective tissues (Fig. 8a). Conversely, the bare sample was compressed upon implantation and continuously lost its reconstructive inner feature. As in Fig. 8b, some new bone (NB) formation is conspicuous in the bare construct, where the scaffold is frequently compressed. Therefore, the effectiveness of the PLGA-coated scaffolds in eliciting bone formation is more than the bare ones.

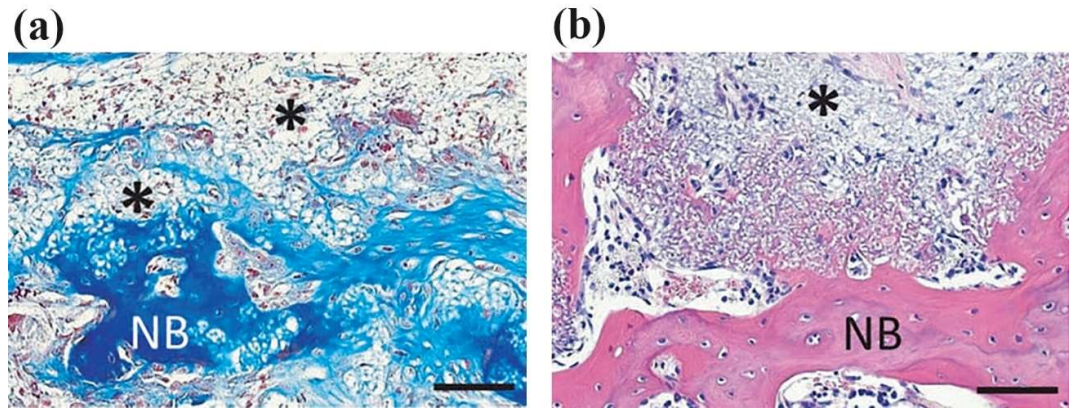

Fig. 8. Histological findings on the bone augmentation of PLGA-coated (a) and bare (b) β-TCP scaffolds (newly formed tissues are indicated by * and the scale bars represent 50 μm), reprinted with permission from Ref. [59].

Zhao et al. [43, 69] evaluated the impact of PLGA coating on the apatite-inducing ability of $CaSiO_3$ scaffolds. It was turned out that when the PLGA-coated scaffolds are immersed in SBF, parts of the $CaSiO_3$ substrate are exposed to SBF along with the degradation of PLGA, providing the nucleation sites for apatite formation. After soaking the samples, apatite nanocrystals were formed on both the bare and coated specimens with a worm-like morphology. Although the polymer coating reduces the pH value during degradation, the apatite formation is not suppressed considerably. The ability of the capsulated scaffolds to encourage apatite precipitation inversely depends on the thickness of the applied coating.

Keihan et al. [61] looked over apatite precipitation on bredigite scaffolds coated with PLGA. Noticeably, the ion-exchange mechanism controls the apatite formation ability of bredigite. It was revealed that the applied PLGA coating slows down the apatite-formation ability kinetics by restricting both bredigite dissolution and the precipitation of ions needed for apatite deposition. Besides, the released acidic products resulting from PLGA biodegradation decline the pH level of the medium, giving rise to the resorption of the precipitated calcium phosphate.

Sadiasa et al. [65] assessed the bioactivity of PLGA/BCP-coated BCP/ZrO$_2$ scaffolds. In this regard, the bare and coated scaffolds were *in vitro* soaked in phosphate-buffered saline (PBS). It was observed that the coating layer is partially deformed during immersion, causing the exposure of the underneath material to the medium and the precipitation of octacalcium phosphate on the samples. This instigates advantageously the osteoconduction, osteoinduction, and thus bone formation of the scaffolds.

The effect of PLGA coating on the osseointegration of titanium implants was monitored by Ravichandran et al. [76]. It was observed that PLGA nanofibers electrospun on the implant surface can enhance the cell adhesion, proliferation, and differentiation, associated to the nanofibers resembling the natural extracellular matrix. *In vitro* outcomes proved that the nanofibrous coating can potentially improve the osseointegration of the Ti implant. Similarly, Hu et al. [36] demonstrated improvements in the angiogenesis and osseointegration of Ti scaffolds after PLGA coating owing to its degradation by-products.

Overall, PLGA coatings limit the scaffold dissolution and precipitation of ions required for apatite precipitation. The degradation of PLGA reduces the pH value of the physiological medium unfavorably, generally leading to a reduction in the apatite formation kinetics with no considerable retarding influence on total bioactivity. From the osseointegration point of view, it can be claimed that the deposited coating is effective in holding the interconnected feature of the scaffold struts prone to be filled with new bone and tissue. But further studies are required *in vivo* to realize the impact of PLGA coatings on the protein adhesion and cellular activities as the second and third steps of bioactivity, respectively, coming after the apatite deposition.

## 6. Conclusions and future research directions

The present article focused on the broad scope of research works conducted on PLGA-coated tissue engineering scaffolds, consisting of substrate materials, coating methods and the influence of the coating on the biological and mechanical capacity. A comprehensive summary of the related information published in the literature is listed in Table 1.

Table 1. Overview of the information published on PLGA coating of various tissue engineering scaffolds

| Substrate materials | Coating Technique | Effect of PLGA coating on | | | | | Reference |
|---|---|---|---|---|---|---|---|
| | | Mechanical properties | Biodegradability | Biocompatibility | Molecular delivery | Bioactivity | |
| Ti | Electrospinning | - | - | Amplifying cell attachment, proliferation, and differentiation | - | Meeting better osseointegration | [76] |
| Ti | Dip-coating | - | - | - | - | Strengthening osteogenesis and osteointegration | [36] |
| HA | Dip-coating and following centrifuging | Upholding mechanical properties upon immersion | Causing a moderate degradation rate | - | - | - | [60] |
| β-TCP | Dip-coating | Eliciting further mechanical strength, reinforcing the scaffold stability | - | Favorable biocompatibility | - | Augmenting bone formation | [59] |
| β-TCP | Dip-coating under vacuum and following centrifuging | Fortifying toughness, compressive and bending strengths | - | Proper cells attachment and proliferation | - | - | [42] |
| β-TCP | Dip-coating and following centrifuging | Increasing compressive strength and toughness | - | Minor influence | Introducing desirable delivery behavior | - | [57] |
| 58S bioactive glass | Dip-coating and following centrifuging | Providing proper mechanical behaviors | - | Enhancing cell migration, attachment, and proliferation | - | - | [62] |
| CaSiO$_3$ | Dip-coating under vacuum | Reaching the optimum mechanical properties | Limiting intrinsic rapid biodegradation of the substrate | Encouraging cell growth and proliferation | Making local stable drug release | Generating suitable nucleation sites for apatite formation | [43, 69] |
| Bredigite | Dip-coating and following centrifuging | Improving the compressive strength | Hindering fast degradation | Highest biocompatibility at 10% concentration of PLGA | Altering the drug delivery behavior | - | [9, 44] |
| Bredigite | Dip-coating | Giving strength close to that of trabecular bone | Controlling the degradation rate | Supporting cellular activities | - | Slowing down apatite-formation kinetics | [61] |
| Bredigite, diopside, akermanite | Dip-coating | - | - | Enhancing biocompatibility | Restricting primary burst drug release | - | [71] |
| PCL | Electrospinning | No significant effect on compressive strength and Young's modulus | - | Sufficient biocompatibility | - | - | [45] |
| Silk | Electrospinning | Inducing superior mechanical characteristic | - | Yielding better cell proliferation | Sustained drug release | - | [63, 64] |
| HA/TCP | Dip-coating and following centrifuging | Providing tougher and stronger structures, the need for improving the mechanical properties further to meet those of human cancellous bone | - | Adequate cell viability | - | - | [66] |
| BCP/ZrO$_2$ | Dip-coating | Advantagous mechanical behaviors | Reducing the weight loss | Making appropriate substrates for cells proliferation | Controlling the primary burst drug release | Modifying calcium phosphate formation on the surface | [65] |

Among coating techniques, dip-coating has been utilized much more by virtue of its simplicity and versatility. Coating and infiltrating of scaffolds with PLGA, as reviewed in the current paper, act as an attractive approach to developing the bio-performance of tissue engineering scaffolds. Accordingly, not only PLGA-coated scaffolds exhibit adjustable degradation kinetics, making them promising for specific-patient applications, but they also ensure cellular activities for more prolonged periods. Other advantages of introducing the PLGA phase as a coating layer on scaffolds are, first, reaching a sustained delivery mode from the scaffolds and, second, getting mechanical characteristics ameliorated. Despite the aforementioned significant properties, this polymer coating almost often adversely affects apatite formation. Despite the published information, there are still areas related to PLGA-coated scaffolds for further investigations in the future, including:

i) The impact of different coating process parameters on the behavior of PLGA-coated scaffolds: On the one hand, the polymer concentration, solvent composition, number of dip-coating and drying passes, immersion duration, and deposition temperature for dip-coating and on the other hand, temperature, pressure, tip-to-target distance, needle diameter, operating voltage, and flow rate for electrospinning are parameters affecting the characteristic of the coatings.

ii) The coating thickness: The influence of the PLGA coating thickness on the bio-performance of scaffolds has not been looked sufficiently. Despite the improvement in the mechanical properties of the scaffolds by increasing the coating thickness, it not only reduces the porosity required for the transmission of nutrients and wastes but also weakens the adhesion of the coating layer to the substrate. Therefore, future research works should take the coating thickness into consideration to reach a compromise between the structural and functional properties.

iii) The effect of PLGA coating on entire bioactivity: As noted above, the PLGA coatings reduce the kinetics of apatite formation as the primary stage of bioactivity. But there remains the need for more research *in vivo* to realize the influence of PLGA coatings in the next steps of bioactivity, i.e., protein adhesion and cellular activities.

iv) Time-dependent mechanical properties and bio-performance: More *in vivo* experiments should be conducted to enlighten the effect of degradation of both coating and substrate on biocompatibility and mechanical properties under long-term conditions.

v) Combination with other modification techniques: Given other ways used to enhance the functionality of tissue engineering scaffolds like structural modification and ionic substitution, it is anticipated that a combination of these methods alongside the PLGA coating approach offers a hopeful prospect of achieving the optimized biological and mechanical properties simultaneously.

**This is the accepted manuscript (postprint) of the following article:**
A. Maadani, F. Davoodian, E. Salahinejad, *Effects of PLGA coating on biological and mechanical behaviors of tissue engineering scaffolds*, Progress in Organic Coatings, 176 (2023) 107406.
https://doi.org/10.1016/j.porgcoat.2023.107406
[31] P. Gentile, V. Chiono, I. Carmagnola and P. V. Hatton, *An overview of poly(lactic-co-glycolic) acid (PLGA)-based biomaterials for bone tissue engineering*, Int J Mol Sci, vol. 15, pp. 3640-59, Feb 28 2014.

[32] N. Samadi, A. Abbadessa, A. Di Stefano, C. F. van Nostrum, T. Vermonden, S. Rahimian*, et al.*, *The effect of lauryl capping group on protein release and degradation of poly(D,L-lactic-co-glycolic acid) particles*, J Control Release, vol. 172, pp. 436-43, Dec 10 2013.

[33] V. Pavot, M. Berthet, J. Rességuier, S. Legaz, N. Handké, S. C. Gilbert*, et al.*, *Poly(lactic acid) and poly(lactic-co-glycolic acid) particles as versatile carrier platforms for vaccine delivery*, Nanomedicine (Lond), vol. 9, pp. 2703-18, Dec 2014.

[34] F. J. Martínez-Vázquez, P. Miranda, F. Guiberteau and A. Pajares, *Reinforcing bioceramic scaffolds with in situ synthesized ε-polycaprolactone coatings*, Journal of Biomedical Materials Research Part A, vol. 101, pp. 3551-3559, 2013.

[35] F. J. Martínez-Vázquez, F. H. Perera, P. Miranda, A. Pajares and F. Guiberteau, *Improving the compressive strength of bioceramic robocast scaffolds by polymer infiltration*, Acta Biomaterialia, vol. 6, pp. 4361-4368, 2010/11/01/ 2010.

[36] X.-F. Hu, Y.-F. Feng, G. Xiang, W. Lei and L. Wang, *Lactic acid of PLGA coating promotes angiogenesis on the interface between porous titanium and diabetic bone*, Journal of Materials Chemistry B, vol. 6, pp. 2274-2288, 2018.

[37] B. Mojarad Shafiee, R. Torkaman, M. Mahmoudi, R. Emadi and E. karamian, *An improvement in corrosion resistance of 316L AISI coated using PCL-gelatin composite by dip-coating method*, Progress in Organic Coatings, vol. 130, pp. 200-205, 2019/05/01/ 2019.

[38] J. Comyn, *Contact angles and adhesive bonding*, International Journal of Adhesion and Adhesives, vol. 12, pp. 145-149, 1992/07/01/ 1992.

[39] D. Hegemann, *4.09 - Plasma Polymer Deposition and Coatings on Polymers*, in *Comprehensive Materials Processing*, S. Hashmi, G. F. Batalha, C. J. Van Tyne and B. Yilbas, Eds., ed Oxford: Elsevier, 2014, pp. 201-228.

[40] A. J. Sanchez-Herencia, *Water Based Colloidal Processing of Ceramic Laminates*, Key Engineering Materials, vol. 333, pp. 39-48, 2007.

[41] S. Kumari, H. R. Tiyyagura, Y. B. Pottathara, K. K. Sadasivuni, D. Ponnamma, T. E. L. Douglas*, et al.*, *Surface functionalization of chitosan as a coating material for orthopaedic applications: A comprehensive review*, Carbohydrate Polymers, vol. 255, p. 117487, 2021/03/01/ 2021.

[42] Y. Kang, A. Scully, D. A. Young, S. Kim, H. Tsao, M. Sen*, et al.*, *Enhanced mechanical performance and biological evaluation of a PLGA coated β-TCP composite scaffold for load-bearing applications*, European Polymer Journal, vol. 47, pp. 1569-1577, 2011/08/01/ 2011.

[43] L. Zhao, C. Wu, K. Lin and J. Chang, *The effect of poly(lactic-co-glycolic acid) (PLGA) coating on the mechanical, biodegradable, bioactive properties and drug release of porous calcium silicate scaffolds*, Biomed Mater Eng, vol. 22, pp. 289-300, 2012.

[44] A. Jadidi, E. Salahinejad, E. Sharifi and L. Tayebi, *Drug-delivery Ca-Mg silicate scaffolds encapsulated in PLGA*, International Journal of Pharmaceutics, vol. 589, p. 119855, 2020/11/15/ 2020.

[45] N. Maurmann, D. P. Pereira, D. Burguez, F. D. A. de S Pereira, P. Inforçatti Neto, R. A. Rezende*, et al.*, *Mesenchymal stem cells cultivated on scaffolds formed by 3D printed PCL matrices, coated with PLGA electrospun nanofibers for use in tissue engineering*, Biomedical Physics & Engineering Express, vol. 3, p. 045005, 2017/06/27 2017.

[46] B. Wang, Y. Wang, T. Yin and Q. Yu, *APPLICATIONS OF ELECTROSPINNING TECHNIQUE IN DRUG DELIVERY*, Chemical Engineering Communications, vol. 197, pp. 1315-1338, 2010/06/18 2010.
34